\documentclass[preprint,showpacs,preprintnumbers,amsmath,amssymb,prl]{revtex4}

\usepackage{graphicx}% Include figure files
\usepackage{dcolumn}% Align table columns on decimal point
\usepackage{bm}% bold math

\begin{document}

%\preprint{APS/123-QED}
% Force line breaks with \\
\title{\emph{In situ}\/\ measurement of three-dimensional ion densities in focused femtosecond pulses}

\author{J. Strohaber}
\email{jstroha1@bigred.unl.edu}
\author{C.J.G.J. Uiterwaal}
\email{cuiterwaal2@unl.edu}
\affiliation{Behlen Laboratory,
Department of Physics and Astronomy, \\The University of Nebraska -
Lincoln, City Campus, Lincoln, NE\ 68588-0111, USA}

\date{\today}% It is always \today, today,

\begin{abstract}
We image spatial distributions of Xe$^{q+}$ ions in the focus of a laser beam of ultrashort, intense pulses in all three dimensions, with a resolution of $\sim$3 $\mu$m and $\sim$12 $\mu$m in the two transverse directions. This allows for studying ionization processes without spatially averaging ion yields. Our \emph{in situ}\/\ ion imaging is also useful to analyze focal intensity profiles and to investigate the transverse modal purity of tightly focused beams of complex light. As an example, the intensity profile of a Hermite-Gaussian beam mode HG$_{1,0}$ recorded with ions is found to be in good agreement with optical images.
\end{abstract}

% PACS, the Physics and Astronomy Classification Scheme.

% 32.80.Rm Multiphoton ionization and excitation to highly excited    states (e.g., Rydberg states)
% 32.80.Wr Other multiphoton processes
% 42.30.Wb Image reconstruction; tomography
% 41.85.Ew Beam profile, beam intensity


\maketitle

After the development of chirped pulse amplification\cite{pess87} in the late 1980s and the discovery of self-mode-locking\cite{spen91} in 1991, laser sources emitting femtosecond-duration pulses have become
routinely available. Laboratories around the world have used such
sources for extensive studies of intense-field processes such as
multiphoton ionization, tunneling ionization, and
above-threshold ionization of atoms and of molecules\cite{post04}. The required atomic electric field of about $10^9$ V/cm or more is
created by focusing the ultrashort pulses in a target gas. Currently, much attention is geared toward the experimental realization and characterization of attosecond pulses using intense-field processes\cite{scri04,scri06}. With such extremely short pulses the dynamics of electrons in atoms and molecules can be followed on their own natural timescale; this now emerging field is called 'attophysics'\cite{krau01}.

A direct consequence of carrying out intense-field experiments with focused laser light is the presence of a broad range of peak intensities across the focal region. To understand some of the experimental consequences of this let us consider a process for which the probability to produce a certain product particle is $P(I)$ when the pulse has peak intensity $I$. Following common practice, we might study the process by adjusting the laser pulse energy such that the highest peak intensity found in the focus equals $I_0$ and then collecting product particles from all over the focal region. Doing so, we obtain a volume-integrated product yield $Y(I_0)$ given by
\begin{equation}
Y(I_0)\propto\int_0^{I_0}P(I)\left|\frac{dV(I,I_0)}{dI}\right|dI,
\label{integrateI2}
\end{equation}
in which $V(I,I_0)$ is the volume in which the local peak intensity
exceeds some value $I$\cite{walk98}. Because of the volumetric weighting in
Eq.~(\ref{integrateI2}), the measured yield $Y(I_0)$ differs from the probability $P(I)$ under investigation. For instance, when $I_0$ is
increased, $Y(I_0)$ ultimately becomes proportional to $I_0^{3/2}$
once the saturation intensity (for which $P(I_0)\rightarrow 1)$ is
exceeded\cite{spei76}. This masks the decrease of $P(I)$ that is often expected for intensities greater than the saturation intensity due to competing higher-order processes. Also, subtle features in $P(I)$ \cite{jone95,tale96} tend to be washed out by volume integration.

In this Letter, we present spatially resolved images of 3-D densities of ion charge states as they are created in focused ultrashort pulses. Thus, what we have realized for use in intense-field ionization research is a \emph{photodynamical test tube}, i.e.\ a volume of known, $\mu$m-sized dimensions exposed to peak intensities that are essentially constant across the volume.
This allows measuring $P(I)$ without spatial integration: one simply defines a sufficiently limited spatial region of interest in the ion image and determines the yields originating from this region.

Another goal of our investigations is to analyze focal intensity distributions. These can be inferred because volume elements having the same peak intensity contain identical ion charge state distributions. The target gas then acts as an \emph{in situ}\/\ sensor of local intensity. In this Letter we analyze images, made with Xe ions, of the transverse profile of an intense, ultrashort focused Hermite-Gaussian HG$_{0,1}$ transverse beam mode. The HG modes are related to the Laguerre-Gaussian modes LG$_{p,\ell}$\cite{sieg86}. We are particularly interested in LG modes because they carry a sharply defined amount of optical orbital angular momentum (OAM)\cite{alle03}, and we wish to study the effect of this quantity on intense-field processes. For this, we need modally pure LG modes. \emph{In situ}\/\ ion imaging will be helpful to assess the degree of modal purity: modal interferences due to contaminating mode components (which have different Gouy phases\cite{sieg86}) should show up in the ion image.

Other techniques have been reported to image foci and/or circumvent volumetric weighting.
First, the intensity-selective scanning (ISS) technique\cite{hans96,walk98,bane99,robs05,brya06} utilizes a narrow slit perpendicular to the laser propagation direction, so ion yields are still integrated over two dimensions. To obtain yields free of volume integration raw ISS data must be deconvolved, which adds noise and only works for Gaussian intensity distributions.
Second, the intensity-difference spectrum (IDS) technique\cite{beni04,wang05} has been shown to work only for a two-dimensional Gaussian profile, and is inherently prone to statistical noise.
Lastly, a number of time-of-flight techniques allow ions created at different locations to arrive at different times. \cite{witz98a,witz98b,jone95,bred04}. The technique of \cite{witz98a} was demonstrated to work well for $\sim$200 $\mu$m resolution\cite{witz98b}, but more systematic investigations are required\cite{witz98a} to explore its applicability for resolutions of a few micrometer. The setup of \cite{bred04} only achieved a spatial resolution of 0.68 mm, and requires cold targets. A focus with a waist of a few mm was dissected into 12 adjacent detection volumes of a few 100 $\mu$m.\cite{jone95}

Compared to the previous methods, the present method achieves a better spatial resolution ($\sim$3 $\mu$m and $\sim$12 $\mu$m in the two transverse directions of the beam). It does \emph{not} require \emph{a priori} knowledge about the intensity distribution, and is \emph{not} limited to Gaussian profiles. We show the first spatially resolved ion distributions in three dimensions with resolution of a few micrometer.

\begin{figure}  % Requires \usepackage{graphicx}
\includegraphics[width=\columnwidth]{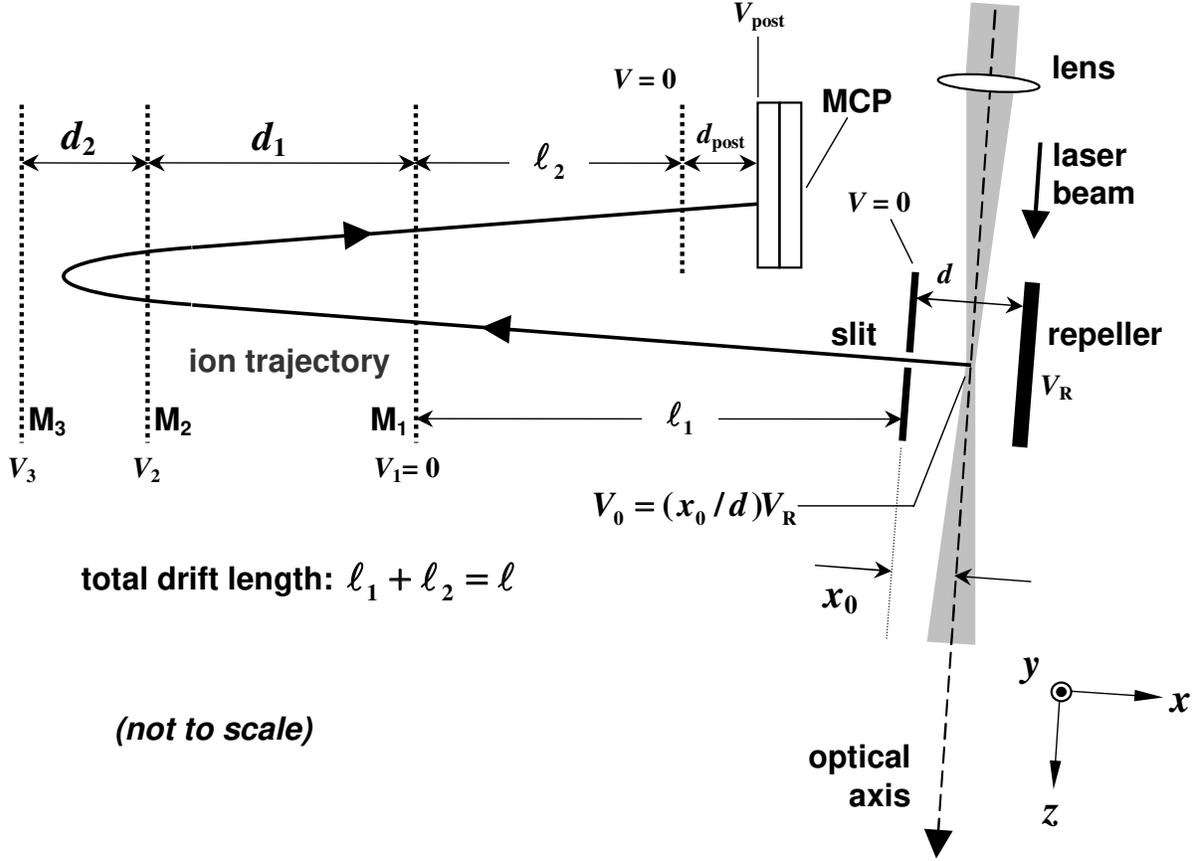}\\
\caption{Schematic of our setup (see text for details).}
\label{fig:setup}
\end{figure}
A schematic of our setup is shown in Fig.\ \ref{fig:setup}. We used an $800$-nm Ti:sapphire laser system (Spectra Physics Tsunami/Spitfire) delivering $\sim$50-fs pulses at a 1-kHz repetition rate with an average output power of about 1.2 W\@. We focus the radiation into an evacuated ionization chamber using an AR-coated UV-grade fused silica plano-convex lens with a nominal focal length of 22.7 cm for 800 nm.
% this is PLCX 38.1-103.0-UV from CVI, f=22.9 cm @ 1064 nm. they give the refractive index of UV-quality fused silica and they (and we) calculate the nominal f = R/(n-1), with R=10.3 cm. note focal spot size estimated by f * lambda / D = 0.229 m * 0.8 um / 0.003(?) m = approx. 60 um, not bad
We lowered the background pressure in the ionization chamber to below $10^{-9}$ mbar using a liquid-N$_2$ trap. We then admitted Xe gas (purity 99.999\%) to a total pressure of $\sim5.0\times10^{-7}$ mbar.
To detect the Xe ions created in the focus, we use a reflectron-type {\sc tof} ion mass spectrometer.
The focus is positioned between two parallel plates. One of these is at ground potential and contains the spectrometer entrance slit; the other plate ('repeller') carries a positive voltage $V_{\rm R}$ of about +1.5 kV and is at a distance $d$ of about 3 mm away. The uniform electrostatic field between the plates causes target gas cations created in the focus to be accelerated toward and pass through the slit. The slit is rectangular, measuring (see coordinate system in Fig.\ \ref{fig:setup}) $\Delta z\sim$400 $\mu$m by $\Delta y\sim$12 $\mu$m. Assuming their initial kinetic energy is negligible, ions created at a distance $x$ away from the slit enter the {\sc tof} tube with kinetic energy proportional to $(x/d)V_{\rm R}$. The total {\sc tof} of an ion with mass $m=Mu$ ($u$ is 1 a.m.u.\ $\approx 1.67\times10^{-27}$ kg) and charge $q=Qe$ ($e$ is the elementary charge $\approx1.60\times10^{-19}$ C) equals
\begin{equation}
t=\sqrt\frac{M}{Q}\ \tau(x).
\label{eq:TOFfunction}
\end{equation}
In this equation, $\tau(x)$ is a reference {\sc tof} of an ion with $M=Q$ (e.g.\ H$^+$) that was created at position $x$; this time depends on the (adjustable) distance $d$ and the fixed distances $d_1$, $d_2$, and $\ell=\ell_1+\ell_2$ between the meshes M$_1$, M$_2$, M$_3$ and the slit, and on the electrostatic potentials $V_2$ and $V_3$ we apply to the meshes (see Fig.\ \ref{fig:setup}; note that $V_1=0$ V). The ion trajectories inside the reflectron have a turning point, and the ions are detected on a multi-channel-plate ion detector ({\sc mcp} in Fig.\ \ref{fig:setup}). To improve the detection efficiency the ions are post-accelerated by applying a potential $V_{\rm post}\sim 8$ kV over a distance $d_{\rm post}$ before they hit the {\sc mcp}. The {\sc mcp} was a Galileo chevron type with a time resolution of $\Delta t\approx 1.5$ ns. To record the ion signal from the {\sc mcp} we used a FAST ComTec model P7886 2-GHz counting card with time bins of 500 ps.

To reconstruct three-dimensional ion distributions, we note that the entrance slit already clips the acceptance volume in the two dimensions $y$ and $z$ to the sizes $\Delta y$ and $\Delta z$ of the slit. We reconstruct the initial positions of ions in the third dimension ($x$) from their {\sc tof} spectrum (Eq.\ \ref{eq:TOFfunction}). To obtain a one-to-one correspondence between initial location $x$ of an ion and its {\sc tof} for each $M/Q$ ratio, we adjust the mesh potentials $V_2$ and $V_3$ so that a monotonically increasing $\tau(x)$ results. For a fixed accumulation time, the signal $S_j$ recorded in time bin $j$ is proportional to the number of ions with {\sc tof} between $t_j = j\Delta t$ and $t_{j+1} =(j+1)\Delta t$ ($\Delta t$ is time bin width). These ions must all originate from a small spatial interval $\Delta x_j=x_{j+1}-x_j$, with $t_j=\sqrt{M/Q}\ \tau(x_j)$ and $t_{j+1}=\sqrt{M/Q}\ \tau(x_{j+1})$. The density of ions $N_j$ created in the spatial interval $\Delta x_j$ is thus
\begin{equation}
N_j\propto\frac{\Delta t_j}{\Delta x_j}
S_j \propto\frac{S_j}{\Delta x_j}.
\label{densities}
\end{equation}
The second proportionality follows because all $\Delta t_j$ are equal. We may think of the recorded {\sc tof} spectrum $S_j$ as the \emph{image in time} of the density $N_j$ of ions in space.
\begin{figure}[t]  % Requires \usepackage{graphicx}
\includegraphics[width=0.8\columnwidth]{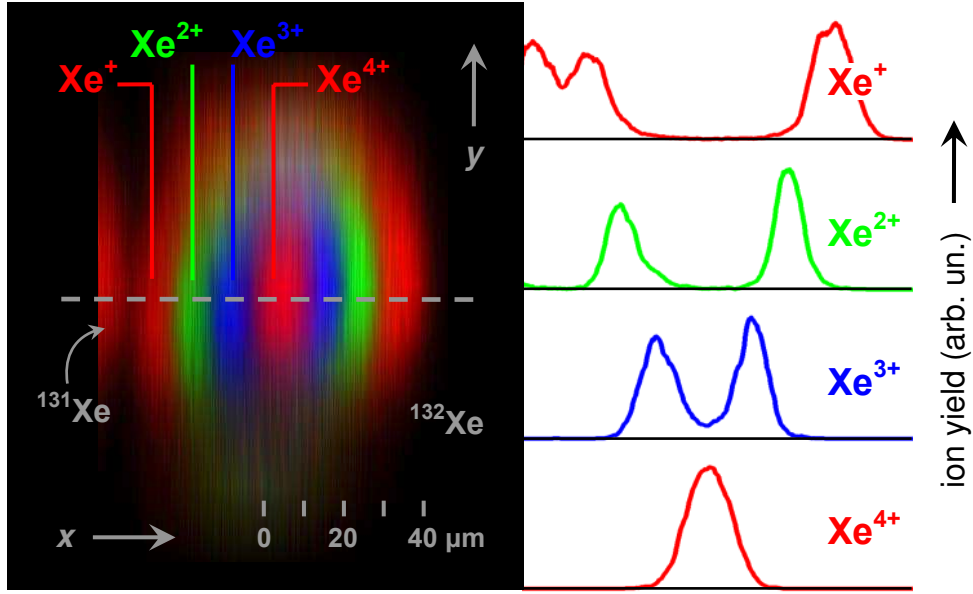}\\
\caption{(Color online) Measured distribution of $^{132}$Xe ions in an $xy$ plane through the focus, with Xe$^+$ in red, Xe$^{2+}$ in green, Xe$^{3+}$ in blue, and Xe$^{4+}$ in red, as indicated. The isotope 132 was used because it is the most abundant. The signal on the left is due to $^{131}$Xe$^+$. Cross sections along the dashed line are shown on the right. The observed nesting of the charge states occurs because the peak intensity is higher at locations closer to the center of the focus. A charge state that dominates at lower-intensity locations is depleted in favor of the next-higher charge state at higher-intensity locations.}
\label{fig:XYscan}
\end{figure}
%Theoretically, this resolution is better when a device with smaller channel widths $\Delta t$ is used. It also improves when we adjust $V_{\rm R}$, $V_1$, $V_2$, and $d$ for a steeper derivative $\left(\partial\tau / \partial x\right)_{x=x_j}$. This latter approach has its limits, though, because it also broadens the {\sc tof} spectrum of each $M/Q$ ratio ion, and for meaningful reconstruction of spatial information we must avoid overlap between two adjacent $M/Q$ ratios. Finally, the resolution depends on the $M/Q$ ratio of the ion under investigation, and is$\sqrt{132/4}\approx 5.7$ times better ($\Delta x$ is smaller) for $^{132}{\rm Xe}^+$ than for $^{4}{\rm He}^+$.
%We used the potentials $V_2\sim825.0$ V and $V_3\sim1048$ V, and positioned the focus in the range 1.9 mm $<x<$ 2.0 mm.
The spatial resolution (in the $x$ direction) is
\begin{equation}
\Delta x_j \approx
\sqrt\frac{Q}{M}\cdot\frac{\Delta t}
{\left(\partial\tau / \partial x\right)_{x=x_j}}
\label{eq:resolution}
\end{equation}
For the typical settings we used, the derivative $\partial\tau/\partial x$ smoothly drops from about 0.250 ns/$\mu$m (at $x=1.9$ mm) to
%0.253 ns/$\mu$m
%0.198 ns/$\mu$m
about 0.200 ns/$\mu$m (at $x=2.0$ mm). To measure $\tau (x)$ we used a focused attenuated beam, which generates only Xe$^+$. We positioned this focus at various $x$ positions using a translation stage with $\mu$m precision, and recorded the {\sc tof} of the Xe$^+$. The obtained experimental $\tau (x)$ matched the theoretically predicted curve. Using Eq.~(\ref{eq:resolution}) with a time resolution $\Delta t = 1.5\ {\rm ns}$ as set by our {\sc mcp}, we find our theoretical spatial resolution for $^{132}{\rm Xe}^+$ ranges between 0.52 $\mu$m (at 1.9 mm) and 0.66 $\mu$m (at 2.0 mm). For $^{132}{\rm Xe}^{7+}$, these values would be 1.4 $\mu$m and 1.7 $\mu$m, respectively. A more realistic estimation of the spatial resolution accounts for the thermal motion of the Xe gas, the dimensions of the spectrometer entrance slit, mechanical vibrations, and the frequency bandwidth of our detection electronics. A first analysis of these factors indicates the spatial resolution in the $x$-direction is about 3 $\mu$m. In the $y$ and $z$ directions the resolutions are mainly determined by the slit dimensions, so 12 $\mu$m and 400 $\mu$m, respectively. Slits narrower than 12 $\mu$m could be used, but the resolution in the $y$ dimension is ultimately limited by thermal effects, in particular for lower charge states. We chose the 400-$\mu$m slit length as a compromise between count rate and the need to keep the slit shorter than the Rayleigh range of our focused beam, which is a few millimeter. Pressure-dependent measurements indicated that space charge effects are absent for the pressure we used ($\sim5.0\times10^{-7}$ mbar).

\begin{figure}[t]  % Requires \usepackage{graphicx}
\includegraphics[width=\columnwidth]{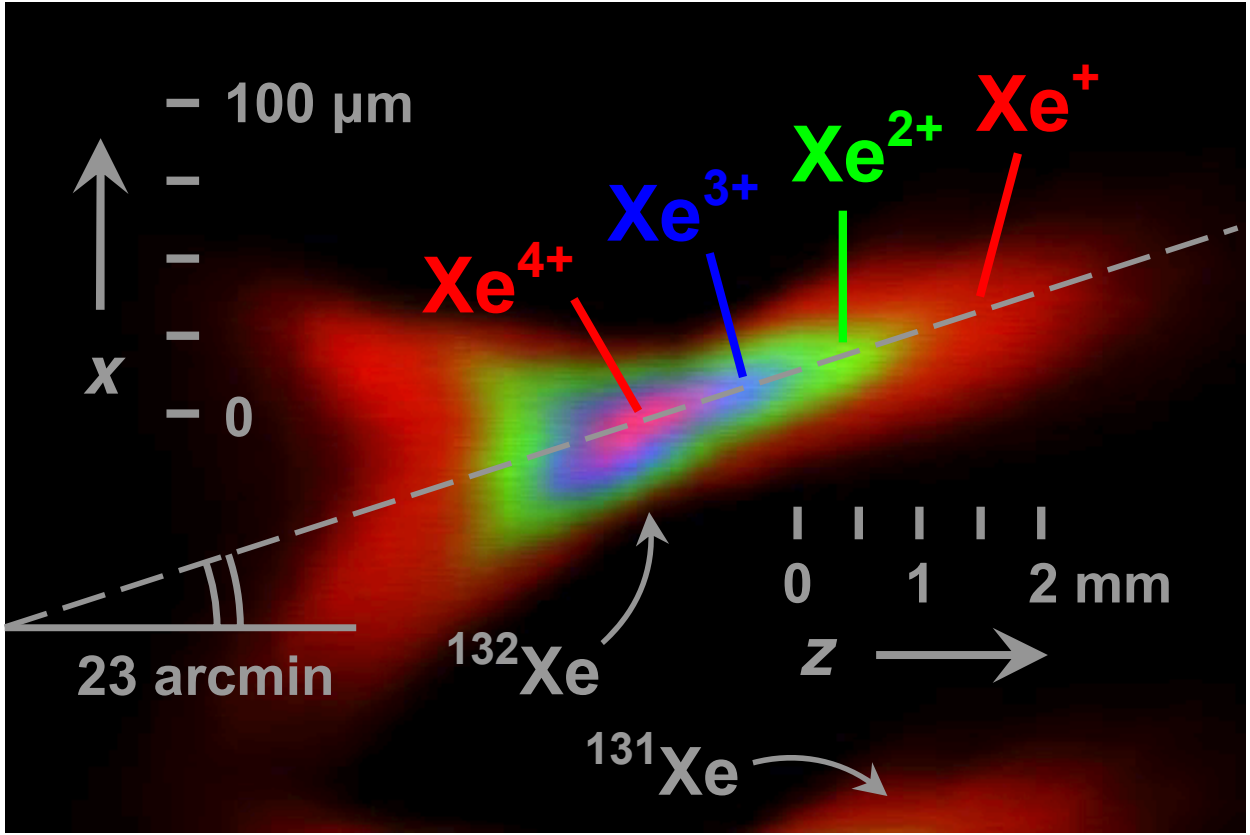}\\
\caption{(Color online) Measured distribution of $^{132}$Xe ions in the $xz$ plane, with Xe$^+$ in red, Xe$^{2+}$ in green, Xe$^{3+}$ in blue, and Xe$^{4+}$ in red, as indicated. The dashed line is the optical axis (beam propagates to the right). Note the difference in scale between the $x$ and $z$ axes. The small angle of 23 arcmin between the optical axis and the $z$ axis is due to a slight misalignment of our setup. The signal at the bottom of the image is due to $^{131}$Xe ions. The observed nesting of the charge states occurs because the peak intensity is higher at locations closer to the center of the focus. A charge state that dominates at lower-intensity locations is depleted in favor of the next-higher charge state at higher-intensity locations. To avoid overlap between adjacent isotopes we used slightly different voltages than those used to obtain Fig.\ \ref{fig:XYscan}}
\label{fig:XZscan}
\end{figure}
Figures \ref{fig:XYscan} and \ref{fig:XZscan} show $^{132}$Xe ion distributions in the focus reconstructed using Eqs.\ \ref{densities} and \ref{eq:resolution}. The different charge states are color-coded: Xe$^+$ (red), Xe$^{2+}$ (green), Xe$^{3+}$ (blue), and Xe$^{4+}$ (red). The image in Fig.\ \ref{fig:XYscan} is a cross section in the $xy$ plane through the focal region, obtained by moving the focus up ($y$ direction) in steps of 10 $\mu$m using a stepper motor, and recording a {\sc tof} spectrum for each height. The image in Fig.\ \ref{fig:XZscan} is a cross section in the $xz$ plane, obtained by moving the focus along the $z$ direction in steps of 200 $\mu$m. Setting up the experiment, we noticed that visible imperfections in the unfocused beam profile go hand in hand with irregular features in the ion charge state distributions. Using this as a guideline, we improved the mode quality of the laser by reducing the aperture of an iris in the regenerative cavity of the amplifier, at the expense of $\sim$0.1 W\@. The nesting of the various charge states in Figs.\ \ref{fig:XYscan} and \ref{fig:XZscan} indicates the competition between the processes that create them. The four line graphs in Fig.\ \ref{fig:XYscan} show the ion yields measured along the dashed line in Fig.\ \ref{fig:XYscan} (transverse to the laser propagation, same scale). If one moves toward the center of the focus, the yield of each charge state Xe$^{q+}$ rises, saturates, and is then depleted in favor of the next charge state Xe$^{(q+1)+}$ in a sequential way, so that a nested pattern of charge states results. The depletion shown in Fig.\ \ref{fig:XYscan} is in sharp contrast with the 3/2-power rise observed in traditional methods. From the data shown in the line graphs, we measured that the count rate of Xe$^+$ and Xe$^{2+}$ in the center of the focus is only a few percent of that of the peaks. Such a degree of observed depletion has no precedence in literature, and underlines our ability to record $P(I)$ free of volumetric weighting.
% http://www.webelements.com/webelements/elements/text/Xe/isot.html gives Xe isotope information

%For alignment purposes, we rigidly mounted two irises to the lens holder, centering them on the lens's optical axis; one iris adjoined the lens and the other one was placed 10 cm in front of it. This lens/iris assembly was mounted on an XYZ motorized translational stage which allowed for precise positioning, with $\sim$2.5-$\mu$m resolution, of the focus in front of the slit. A digital micrometer with a 1-$\mu$m resolution (Fowler Sylvac Ultra Digit Mark IV) was used as a feedback for the positioning of the translational stage.
% A {\sc pc} running a LabVIEW program was used to vertically move the focusing lens in 10-$\mu$m steps allowing us to scan the focus across the slit.

\begin{figure}[t]  % Requires \usepackage{graphicx}  was 92%
\includegraphics[width=\columnwidth]{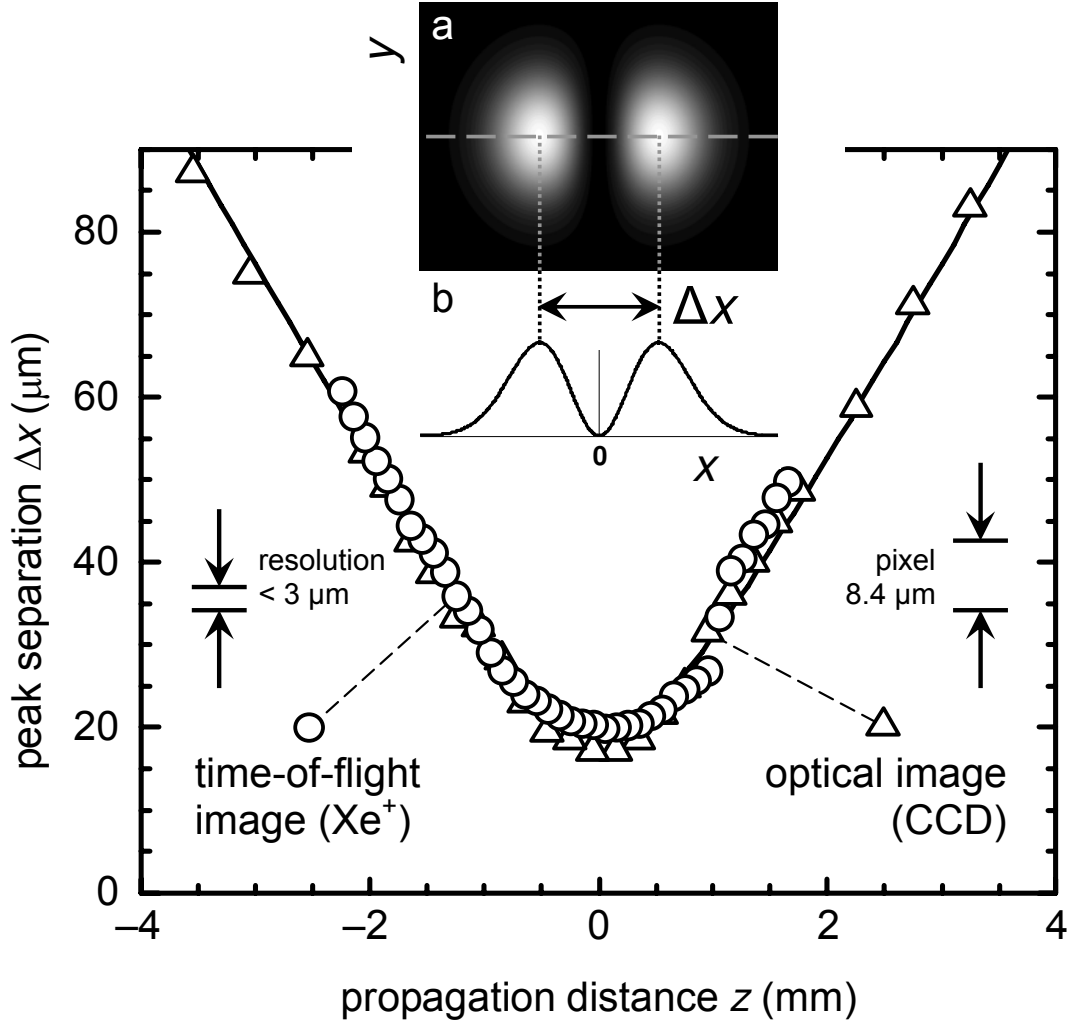}\\
\caption{Focused HG$_{1,0}$ beam: distance $\Delta x$ between the two intensity lobes (see insets) as a function of the propagation distance $z$. Circles: from Xe$^+$ image (resolution $< 3\ \mu$m); triangles: from CCD image (pixel size $8.4\ \mu$m). Solid curve: fit to Eq.\ (\ref{eq:HGPeakSep}).}
\label{fig:HG10}
\end{figure}

To demonstrate that our ion imaging technique is not limited to only Gaussian intensity profiles, we also imaged a focused beam of ultrashort pulses in the Hermite-Gaussian HG$_{1,0}$ transverse mode\cite{sieg86} (Fig.\ \ref{fig:HG10}). We created this mode using holographic techniques\cite{mari05,stro07b}. The intensity of the HG$_{1,0}$ mode is
\begin{equation}
I(x,y,z) \propto \frac{x^2}{w(z)^2} \exp \left(-\frac{x^2+y^2}{w(z)^2}\right),
\label{eq:HGfield}
\end{equation}
in which $w(z)^2=w_0^2\left( 1+z^2/z_0^2\right)$ describes the self-similar transverse scaling of the intensity profile ($w_0$ is the beam waist and $z_0$ is the Rayleigh range)\cite{sieg86}. The profile consists of two lobes (see inset $a$ in Fig.\ \ref{fig:HG10}). For $y=0$ (dashed line in inset) the maxima of these lobes are separated in the $x$ direction by a distance
\begin{equation}
\Delta x(z)=\Delta x_0 \sqrt{1+z^2/z_0^2}
\label{eq:HGPeakSep}
\end{equation}
(see inset $b$), with $\Delta x_0$ the separation at the waist ($z$=0). Again using Eqs.\ \ref{densities} and \ref{eq:resolution}, we reconstructed $\Delta x(z)$ (circles in Fig.\ \ref{fig:HG10}) from {\sc tof} spectra.

To verify that this reconstruction is correct we also focused the beam in air using the same lens as in the ionization experiment, and recorded the beam profile as a function of $z$ optically (triangles in Fig.\ \ref{fig:HG10}) using a CCD camera. In contrast to the ion imaging experiments---taken at full power---this optical experiment required substantial attenuation, to protect the camera from damage and to avoid nonlinear optical effects in air. The ion image and optical image agree within the error limits, which we estimate to be $\sim$8.4 $\mu$m (pixel size) for the CCD data and $\sim$3 $\mu$m for the ion data. This agreement allows a fit to Eq.~(\ref{eq:HGPeakSep}) (solid curve in Fig.\ \ref{fig:HG10}).

In summary, we used a time-of-flight technique to image ion distributions in the focus of a laser beam of ultrafast pulses. We showed that this (\emph{a}) allows measurements of ion yields without volumetric weighting, and (\emph{b}) facilitates investigations of focal intensity distributions, with the target atoms acting as \emph{in situ}\/\ intensity sensors.

This material is based upon work supported by the National Science Foundation under Grant No.\ PHY-0355235. The authors gratefully acknowledge the Max-Planck-Institute of Quantum Optics in Germany, in particular Dr.\ Hartmut Schr\"oder, for generously lending us the reflectron used in this work.


\begin{thebibliography}{23}
\expandafter\ifx\csname natexlab\endcsname\relax\def\natexlab#1{#1}\fi
\expandafter\ifx\csname bibnamefont\endcsname\relax
  \def\bibnamefont#1{#1}\fi
\expandafter\ifx\csname bibfnamefont\endcsname\relax
  \def\bibfnamefont#1{#1}\fi
\expandafter\ifx\csname citenamefont\endcsname\relax
  \def\citenamefont#1{#1}\fi
\expandafter\ifx\csname url\endcsname\relax
  \def\url#1{\texttt{#1}}\fi
\expandafter\ifx\csname urlprefix\endcsname\relax\def\urlprefix{URL }\fi
\providecommand{\bibinfo}[2]{#2}
\providecommand{\eprint}[2][]{\url{#2}}

\bibitem[{\citenamefont{Pessot et~al.}(1987)\citenamefont{Pessot, Maine, and
  Mourou}}]{pess87}
\bibinfo{author}{\bibfnamefont{M.}~\bibnamefont{Pessot}},
  \bibinfo{author}{\bibfnamefont{P.}~\bibnamefont{Maine}}, \bibnamefont{and}
  \bibinfo{author}{\bibfnamefont{G.}~\bibnamefont{Mourou}},
  \bibinfo{journal}{Opt.\ Comm.} \textbf{\bibinfo{volume}{62}},
  \bibinfo{pages}{419} (\bibinfo{year}{1987}).

\bibitem[{\citenamefont{Spence et~al.}(1991)\citenamefont{Spence, Kean, and
  Sibbett}}]{spen91}
\bibinfo{author}{\bibfnamefont{D.~E.} \bibnamefont{Spence}},
  \bibinfo{author}{\bibfnamefont{P.~N.} \bibnamefont{Kean}}, \bibnamefont{and}
  \bibinfo{author}{\bibfnamefont{W.}~\bibnamefont{Sibbett}},
  \bibinfo{journal}{Opt.\ Lett.} \textbf{\bibinfo{volume}{16}},
  \bibinfo{pages}{42} (\bibinfo{year}{1991}).

\bibitem[{\citenamefont{Posthumus}(2004)}]{post04}
\bibinfo{author}{\bibfnamefont{J.~H.} \bibnamefont{Posthumus}},
  \bibinfo{journal}{Rep.\ Progr.\ Phys.} \textbf{\bibinfo{volume}{67}},
  \bibinfo{pages}{623} (\bibinfo{year}{2004}).

\bibitem[{\citenamefont{Scrinzi and Muller}(2004)}]{scri04}
\bibinfo{author}{\bibfnamefont{A.}~\bibnamefont{Scrinzi}} \bibnamefont{and}
  \bibinfo{author}{\bibfnamefont{H.~G.} \bibnamefont{Muller}},
  \emph{\bibinfo{title}{Strong Field Physics}} (\bibinfo{publisher}{Springer},
  \bibinfo{year}{2004}), chap. \bibinfo{chapter}{Attosecond pulses: generation,
  detection, and applications}.

\bibitem[{\citenamefont{Scrinzi et~al.}(2006)\citenamefont{Scrinzi, {Yu.\
  Ivanov}, Kienberger, and Villeneuve}}]{scri06}
\bibinfo{author}{\bibfnamefont{A.}~\bibnamefont{Scrinzi \emph{et al.}}},
  \bibinfo{journal}{J. Phys.\ B}
  \textbf{\bibinfo{volume}{39}}, \bibinfo{pages}{R1} (\bibinfo{year}{2006}).

\bibitem[{\citenamefont{Krausz}(2001)}]{krau01}
\bibinfo{author}{\bibfnamefont{F.}~\bibnamefont{Krausz}},
  \bibinfo{journal}{Phys.\ World} \textbf{\bibinfo{volume}{Sept/2001}},
  \bibinfo{pages}{41} (\bibinfo{year}{2001}).

\bibitem[{\citenamefont{Walker et~al.}(1998)\citenamefont{Walker, Hansch, and
  {Van Woerkom}}}]{walk98}
\bibinfo{author}{\bibfnamefont{M.~A.} \bibnamefont{Walker}},
  \bibinfo{author}{\bibfnamefont{P.}~\bibnamefont{Hansch}}, \bibnamefont{and}
  \bibinfo{author}{\bibfnamefont{L.~D.} \bibnamefont{{Van Woerkom}}},
  \bibinfo{journal}{Phys.\ Rev.\ A} \textbf{\bibinfo{volume}{57}},
  \bibinfo{pages}{R701} (\bibinfo{year}{1998}).

\bibitem[{\citenamefont{Speiser and Jortner}(1976)}]{spei76}
\bibinfo{author}{\bibfnamefont{S.}~\bibnamefont{Speiser}} \bibnamefont{and}
  \bibinfo{author}{\bibfnamefont{J.}~\bibnamefont{Jortner}},
  \bibinfo{journal}{Chem.\ Phys.\ Lett.} \textbf{\bibinfo{volume}{44}},
  \bibinfo{pages}{399} (\bibinfo{year}{1976}).

\bibitem[{\citenamefont{Jones}(1995)}]{jone95}
\bibinfo{author}{\bibfnamefont{R.~R.} \bibnamefont{Jones}},
  \bibinfo{journal}{Phys.\ Rev.\ Lett.} \textbf{\bibinfo{volume}{74}},
  \bibinfo{pages}{1091} (\bibinfo{year}{1995}).

\bibitem[{\citenamefont{Talebpour et~al.}(1996)\citenamefont{Talebpour, Chien,
  and Chin}}]{tale96}
\bibinfo{author}{\bibfnamefont{A.}~\bibnamefont{Talebpour}},
  \bibinfo{author}{\bibfnamefont{C.-Y.} \bibnamefont{Chien}}, \bibnamefont{and}
  \bibinfo{author}{\bibfnamefont{S.~L.} \bibnamefont{Chin}},
  \bibinfo{journal}{J. Phys.\ B} \textbf{\bibinfo{volume}{29}},
  \bibinfo{pages}{5725} (\bibinfo{year}{1996}).

\bibitem[{\citenamefont{Siegman}(1986)}]{sieg86}
\bibinfo{author}{\bibfnamefont{A.~E.} \bibnamefont{Siegman}},
  \emph{\bibinfo{title}{Lasers}} (\bibinfo{publisher}{University Science Books,
  Sausalito, CA}, \bibinfo{year}{1986}).

\bibitem[{\citenamefont{Allen et~al.}(2003)\citenamefont{Allen, Barnett, and
  Padgett}}]{alle03}
\bibinfo{editor}{\bibfnamefont{L.}~\bibnamefont{Allen}},
  \bibinfo{editor}{\bibfnamefont{S.~M.} \bibnamefont{Barnett}},
  \bibnamefont{and} \bibinfo{editor}{\bibfnamefont{M.~J.}
  \bibnamefont{Padgett}}, eds., \emph{\bibinfo{title}{Optical angular
  momentum}} (\bibinfo{publisher}{Inst. of Phys. Publishing},
  \bibinfo{year}{2003}).

\bibitem[{\citenamefont{Hansch et~al.}(1996)\citenamefont{Hansch, Walker, and
  {Van Woerkom}}}]{hans96}
\bibinfo{author}{\bibfnamefont{P.}~\bibnamefont{Hansch}},
  \bibinfo{author}{\bibfnamefont{M.~A.} \bibnamefont{Walker}},
  \bibnamefont{and} \bibinfo{author}{\bibfnamefont{L.~D.} \bibnamefont{{Van
  Woerkom}}}, \bibinfo{journal}{Phys.\ Rev.\ A} \textbf{\bibinfo{volume}{54}},
  \bibinfo{pages}{R2559} (\bibinfo{year}{1996}).

\bibitem[{\citenamefont{Banerjee et~al.}(1999)\citenamefont{Banerjee, {Ravindra
  Kumar}, and Mathur}}]{bane99}
\bibinfo{author}{\bibfnamefont{S.}~\bibnamefont{Banerjee}},
  \bibinfo{author}{\bibfnamefont{G.}~\bibnamefont{{Ravindra Kumar}}},
  \bibnamefont{and} \bibinfo{author}{\bibfnamefont{D.}~\bibnamefont{Mathur}},
  \bibinfo{journal}{J. Phys.\ B} \textbf{\bibinfo{volume}{32}},
  \bibinfo{pages}{L305} (\bibinfo{year}{1999}).

\bibitem[{\citenamefont{Robson et~al.}(2005)\citenamefont{Robson, Ledingham,
  McKenna, McCanny, Shimizu, Yang, Wahlstr\"{o}m, Lopez-Martens, Varju,
  Johnsson et~al.}}]{robs05}
\bibinfo{author}{\bibfnamefont{L.}~\bibnamefont{Robson\emph{ et al.}}},
  \bibinfo{journal}{J. Am.\ Soc.\ Mass Spectrom.}
  \textbf{\bibinfo{volume}{16}}, \bibinfo{pages}{82} (\bibinfo{year}{2005}).

\bibitem[{\citenamefont{Bryan et~al.}(2006)\citenamefont{Bryan, Stebbings,
  English, Goodworth, Newell, J~McKenna, Williams, Turcu, Smith, Divall
  et~al.}}]{brya06}
\bibinfo{author}{\bibfnamefont{W.~A.} \bibnamefont{Bryan \emph{et al.}}},
  \bibinfo{journal}{Phys.\ Rev.\
  A} \textbf{\bibinfo{volume}{73}}, \bibinfo{pages}{013407}
  (\bibinfo{year}{2006}).

\bibitem[{\citenamefont{Wang et~al.}(2005)\citenamefont{Wang, Sayler, Carnes,
  Esry, and Ben-Itzhak}}]{wang05}
\bibinfo{author}{\bibfnamefont{P.~Q.} \bibnamefont{Wang \emph{et al.}}},
  \bibinfo{journal}{Opt.\ Lett.} \textbf{\bibinfo{volume}{30}},
  \bibinfo{pages}{664} (\bibinfo{year}{2005}).

\bibitem[{\citenamefont{Benis et~al.}(2004)\citenamefont{Benis, Xia, Tong,
  Faheem, Zamkov, Shan, Richard, and Chang}}]{beni04}
\bibinfo{author}{\bibfnamefont{E.~P.} \bibnamefont{Benis \emph{et al.}}},
  \bibinfo{journal}{Phys.\ Rev.\ A} \textbf{\bibinfo{volume}{70}},
  \bibinfo{pages}{025401} (\bibinfo{year}{2004}).

\bibitem[{\citenamefont{Witzel et~al.}(1998{\natexlab{a}})\citenamefont{Witzel,
  Schr\"{o}der, Kaesdorf, and Kompa}}]{witz98a}
\bibinfo{author}{\bibfnamefont{B.}~\bibnamefont{Witzel \emph{et al.}}},
  \bibinfo{journal}{Int.\ J. Mass Spectrom.\ Ion Proc.}
  \textbf{\bibinfo{volume}{172}}, \bibinfo{pages}{229}
  (\bibinfo{year}{1998}{\natexlab{a}}).

\bibitem[{\citenamefont{Witzel et~al.}(1998{\natexlab{b}})\citenamefont{Witzel,
  Uiterwaal, Schr\"{o}der, Charalambidis, and Kompa}}]{witz98b}
\bibinfo{author}{\bibfnamefont{B.}~\bibnamefont{Witzel \emph{et al.}}},
  \bibinfo{journal}{Phys.\ Rev.\ A} \textbf{\bibinfo{volume}{58}},
  \bibinfo{pages}{3836} (\bibinfo{year}{1998}{\natexlab{b}}).

\bibitem[{\citenamefont{Bredy et~al.}(2004)\citenamefont{Bredy, Camp, Nguyen,
  Awata, Shan, Chang, and DePaola}}]{bred04}
\bibinfo{author}{\bibfnamefont{R.}~\bibnamefont{Bredy \emph{et al.}}},
  \bibinfo{journal}{J. Opt.\ Soc.\ Am.\ B} \textbf{\bibinfo{volume}{21}},
  \bibinfo{pages}{2221} (\bibinfo{year}{2004}).

\bibitem[{\citenamefont{Mariyenko et~al.}(2005)\citenamefont{Mariyenko,
  Strohaber, and Uiterwaal}}]{mari05}
\bibinfo{author}{\bibfnamefont{I.~G.} \bibnamefont{Mariyenko}},
  \bibinfo{author}{\bibfnamefont{J.}~\bibnamefont{Strohaber}},
  \bibnamefont{and} \bibinfo{author}{\bibfnamefont{C.~J. G.~J.}
  \bibnamefont{Uiterwaal}}, \bibinfo{journal}{Opt.\ Express}
  \textbf{\bibinfo{volume}{13}}, \bibinfo{pages}{7599} (\bibinfo{year}{2005}).

\bibitem[{\citenamefont{Strohaber et~al.}(2007)\citenamefont{Strohaber,
  Petersen, and Uiterwaal}}]{stro07b}
\bibinfo{author}{\bibfnamefont{J.}~\bibnamefont{Strohaber}},
  \bibinfo{author}{\bibfnamefont{C.}~\bibnamefont{Petersen}}, \bibnamefont{and}
  \bibinfo{author}{\bibfnamefont{C.~J. G.~J.} \bibnamefont{Uiterwaal}},
  \bibinfo{journal}{Opt.\ Lett.} \textbf{\bibinfo{volume}{32}},
  \bibinfo{pages}{2387} (\bibinfo{year}{2007}).

\end{thebibliography}
\end{document}